\documentclass{article}

\usepackage{arxiv}

\usepackage[utf8]{inputenc} 
\usepackage[T1]{fontenc}    
\usepackage{hyperref}       
\usepackage{url}            
\usepackage{booktabs}       
\usepackage{amsfonts}       
\usepackage{nicefrac}       
\usepackage{microtype}      
\usepackage{lipsum}

\usepackage{cite}
\usepackage{booktabs}
\usepackage{lipsum, mdframed}  
\usepackage{tabularx}

\usepackage[pdftex]{graphicx}

\usepackage{nicefrac}
\newcommand{\rpm}{\sbox0{$1$}\sbox2{$\scriptstyle\pm$}
  \raise\dimexpr(\ht0-\ht2)/2\relax\box2 }
\usepackage{amsmath}
\usepackage[table,x11names]{xcolor}
\definecolor{myblue}{HTML}{009696}
\definecolor{mypurple}{HTML}{5e3c99}
\definecolor{myorange}{HTML}{e66101}

\usepackage{algorithm}
\usepackage{algpseudocode}
\newfloat{algorithm}{t}{lop}

\usepackage{fourier,tikz}
\usetikzlibrary{arrows,calc}

\title{The Ambiguous World of Emotion Representation}

\author{
    Vidhyasaharan Sethu \\
  {\small School of Electrical Engineering and Telecommunications} \\
  {\small University of New South Wales} \\
  {\small Sydney, NSW 2052, Australia} \\
  {\small \texttt{v.sethu@unsw.edu.au}} \\
   \And
 Emily Mower Provost \\
  {\small Department of Electrical Engineering and Computer Science} \\
  {\small University of Michigan} \\
  {\small Ann Arbor, MI48109, USA} \\
  {\small \texttt{emilykmp@umich.edu}} \\
   \AND
   Julien Epps \\
  {\small School of Electrical Engineering and Telecommunications} \\
  {\small University of New South Wales} \\
  {\small Sydney, NSW 2052, Australia} \\
  {\small \texttt{j.epps@unsw.edu.au}} \\
   \And
   {\small Carlos Busso} \\
   {\small Erik Jonsson School of Engineering \& Computer Science} \\
   {\small The University of Texas at Dallas} \\
   {\small Richardson TX 75080, USA} \\
   {\small \texttt{busso@utdallas.edu}} \\
   \And
   {\small Nicholas Cummins} \\
   {\small ZD.B Chair of Embedded Intelligence for Health Care and Wellbeing} \\
   {\small University of Augsburg} \\
   {\small Augsburg, Germany} \\
   {\small \texttt{nicholas.cummins@informatik.uni-augsburg.de}} \\
  \And
   {\small Shrikanth Narayanan} \\
   {\small Signal Analysis and Interpretation Lab} \\
   {\small University of Southern California, Los Angeles} \\
   {\small CA 90089, USA} \\
   {\small \texttt{shri@ee.usc.edu}} \\
}

\begin{document}
\maketitle

\begin{abstract}
Artificial intelligence and machine learning systems have demonstrated huge improvements and human-level parity in a range of activities, including speech recognition, face recognition and speaker verification. However, these diverse tasks share a key commonality that is not true in affective computing: the ground truth information that is inferred can be unambiguously represented. This observation provides some hints as to why affective computing, despite having attracted the attention of researchers for years, may not still be considered a mature field of research. A key reason for this is the lack of a common mathematical framework to describe all the relevant elements of emotion representations. This paper proposes the AMBiguous Emotion Representation (AMBER) framework to address this deficiency. AMBER is a unified framework that explicitly describes categorical, numerical and ordinal representations of emotions, including time varying representations. In addition to explaining the core elements of AMBER, the paper also discusses how some of the commonly employed emotion representation schemes can be viewed through the AMBER framework, and concludes with a discussion of how the proposed framework can be used to reason about current and future affective computing systems.
\end{abstract}


\section{Introduction}
\label{s:intro}

Emotions are an essential part of natural human-computer interaction (e.g. \cite{Calvo10-ADA, Esposito15-NAC, Vinciarelli15-OCI}). There is considerable potential to augment technologies by leveraging information about users' emotions \cite{picard2000affective}, for a range of applications from improving human communication \cite{lee2010quantification} to health care \cite{picard1997affective}. However, current emotion recognition systems use models of emotions that are restricted, compared with those that we humans use, which fundamentally limits their utility. For example, consider the following scene: a user has had a rough day and upon arriving at home sarcastically interacts with a virtual assistant.  The virtual assistant, however, is not equipped for the full range of human sarcasm, and it fails to react ``appropriately.'' By contrast, a human listener, whose mental model of emotions is much richer, would be capable of recognising sarcasm or, if the expression truly is ambiguous, would respond by eliciting more information to help resolve the ambiguity. The essential challenge is that emotions are complicated and current methods of mathematically describing emotions often do not take the full complexity of the signal into account.  This manuscript presents and motivates the \emph{AMBiguous Emotion Representation} (AMBER) framework, which is explicitly designed to describe and reason about emotion representations, including the ambiguity that is a natural part of emotional displays.

Current affective computing systems generally represent emotions with categorical labels (e.g., angry, happy) or as values on numerical scales (e.g., arousal/activation - calm vs. excited,  valence - negative vs. positive) \cite{gunes2013categorical,schroder2007should}. There has also been interest in ordinal representations that acknowledge the inherently ordinal characteristic of emotion perception \cite{martinez2014don, Yannakakis_2017,Yannakakis_201x} (e.g., medium valence on a scale of \{low, medium, high\}). However, all three types of labels are single-valued point estimates that by their very nature cannot quantify the ambiguity present in emotions, both expressed and perceived.

There are several examples of emotion representations that acknowledge this problem. For example, Vidrascu and Devillers proposed the notion of blended emotions, represented using both a major and a minor categorical label \cite{vidrascu2005real}, and a similar two-category approach without the major/minor distinction has also been proposed \cite{sobol2010classification}. The emotion profile representation is a more general approach that describes emotion in terms of the numerical intensity of a finite number of emotion primitives \cite{grimm2007primitives, mower2011framework}, and other authors have similarly noted that discrete emotions with intensities should be considered \cite{alzoubi2012detecting}. The entire notion of annotator agreement, which is pervasive in descriptions of affective datasets, implicitly acknowledges the ambiguity in emotion perception among any group of annotators. This ambiguity is a central feature in emotion recognition, from problem formulation to system evaluation.


Research in emotion has, thus far, naturally incorporated the assumptions from these domains: variability among annotators is  undesirable and should be treated as noise \cite{chao2015long}, averaged to produce a point estimate \cite{schuller2012avec}, or removed to arrive at a dominant emotion label \cite{mower2009evaluating}. However, concepts of annotation noise, averaged annotations or dominant emotions are at best reductionist assumptions that may be very unrealistic in many practical instances.   Emotion representations need to be capable of reflecting the diversity of human annotation, due to the inherently subjective nature of affective experiences, both while expressing and perceiving emotions. 

There are relatively few examples of emotion representations in affective computing to date that can represent this diversity.  Existing efforts include soft labels where a numerical intensity is associated with each emotion category \cite{zhou2015emotion, fayek2016modeling, kim2018human, mower2011framework,Lotfian_2017}, inclusion of emotion distributions \cite{joshi2011aesthetics,wang2015modeling,zhang2017predicting}, and the use of confidence \cite{zovato201529}. Despite this, a single-valued average emotional label has almost always been adopted as the \emph{ground truth} representation, and modelling the ambiguity as expressed by inter-observer agreement levels within automatic affect analysers remains a noted challenge in the field \cite{gunes2011emotion}. The essential challenge is that current emotion representations, and indeed machine learning approaches employed in affective computing in general, lack the capability to deal adequately with subjective quantities such as emotions, where the information to be modelled and inferred may be inherently ambiguous. It should be noted at this stage that we restrict the use of the term \textit{ambiguity} (in this paper) to refer to the challenge arising from subtlety in expression and differences in perception. We also make the distinction between \textit{ambiguity} and \textit{uncertainty}, a term we use to indicate impreciseness of machine learning models trained on a finite dataset. 

This paper explores explicit and implicit assumptions that are inherent to common emotion representation schemes (Section 2) and the ability of such schemes to represent ambiguity and uncertainty (Section 3). It also identifies desirable characteristics of emotion representation schemes, and proposes a new general framework for generating and comparing representations, termed AMBER, which is explicitly designed to accommodate ambiguity and uncertainty in all contexts (Section 4). The paper also shows examples of how AMBER can unify existing emotion representations, reason about them, and expand them to incorporate ambiguity and uncertainty into affective technologies.

\section{Current Uses of Affective Computing}
Emotion representations are widely used in a range of \emph{Artificial Intelligence} (AI) based technologies. 
Example technologies include: affective-aware \emph{Human-Computer Interaction} (HCI) and \emph{Human-Machine Interface} (HMI) frameworks; supportive technologies in remote education and healthcare systems, and intelligent business and customer service systems~\cite{Harley17-ATO,jaimes07-MHC}. These technologies can broken down into two core applications: (1) emotion perception and (2) emotion synthesis and conversion.  The first requires an emotion representation to interpret human emotions. The second requires an emotion representation for the generation of emotion.  In the remainder of this section we provide an overview for existing technologies that use affective computing.

Healthcare is a growing area of affective computing. Affective technologies can aid in diagnosis (e.g., providing objective markers for mental health conditions such as depression~\cite{Cummins15-ARO,Pampouchidou17-AAO}) and symptom severity tracking (e.g., tracking mania and depression symptoms for populations with bipolar disorder~\cite{khorram2018priori, matton2019transitioning}). Affective technologies can also provide the emotional reasoning that enables embedded agents -- virtual or robotic -- to provide support services or in cybertherapy settings~\cite{Bickmore10-RAI,Messinger14-ACE, Riva14-CAA}. 

Humans have a tendency to anthropomorphize computers and related technologies, displaying social attitudes and behaviours towards them. Agents with greater empathic accuracy are, therefore, considered to be more effective at comforting users, even if this improved accuracy is at the cost of restricting user input behaviours~\cite{Bickmore07-PAT}. However, in order to achieve this improved empathic response without restricting user inputs, affective reasoning systems must have access to richer models of emotion that take ambiguity into account. Consider affect-aware learning technologies, technologies that recognise and react to a user's current emotional state to create emotionally supportive learning environments~\cite{Calvo12-FOA,Dmello14-FTA}. The creation of such environments presupposes the ability to operate in naturalistic settings, and adapt learning tasks using fail-soft paradigms, which ensure the learning process is not impaired by incorrectly recognising the user's emotional state~\cite{Dmello14-FTA}.  Fundamentally, they require the ability to characterise the emotional richness, the emotional ambiguity, of the learning interaction.  

\emph{Internet-of-Things} (IoT) applications must also operate in naturalistic and ambiguous environments.  These applications track discrete emotion representations often in a distinct, binary or triadic classification paradigms, such as positive affective state recognition, happiness recognition, and boredom recognition~\cite{Politou17-ASO}. IoT applications generally require computation on mobile platforms, which limits the computational complexity of the algorithms.  However, with increases in the computational power of mobile devices, the feasibility of complex emotion recognition has been greatly enhanced.  

\emph{Affective Gaming} (AG) uses affective measurements to enhance the gaming experience~\cite{Christy14-TAI, Yannakakis14-EIG}. Video games often aim to elicit emotions through gameplay characteristics such as story-lines, characters and their (perceived) personalities, and video and music effects. AG takes this one step further, by aiming to detect the emotional reactions of the player, and use these reactions to adjust gameplay. This task is challenging as gaming generally does not occur in controlled laboratory environments.  Yet, the constraints that are present in gaming offer an opportunity to collect rich affective data that include self-annotation, which can enable semi-supervised and active learning paradigms~\cite{Christy14-TAI}. In this domain, it is less important to know the exact emotional state than in other domains such as affective tutoring,. What matters is emotional changes.  Furthermore, players are more willing to be in negative emotional states -- frustration, fear, anxiety -- if it enhances their overall gaming experience~\cite{Yannakakis14-EIG}.

However, despite the prolific use of affective computing in existing technology, most affective technologies are not designed to handle ambiguity.  Consequently, they cannot reason about, and react appropriately to, affect that is ambiguous. This is critically important.  Affective computing technologies will not be adopted if they cannot correctly reason over the full span of human emotion expression, or if they seek to make concrete judgements about underlying affect when absolute judgement is neither warranted nor possible.

\begin{figure}
 \begin{center}
    \includegraphics[width=0.4\textwidth]{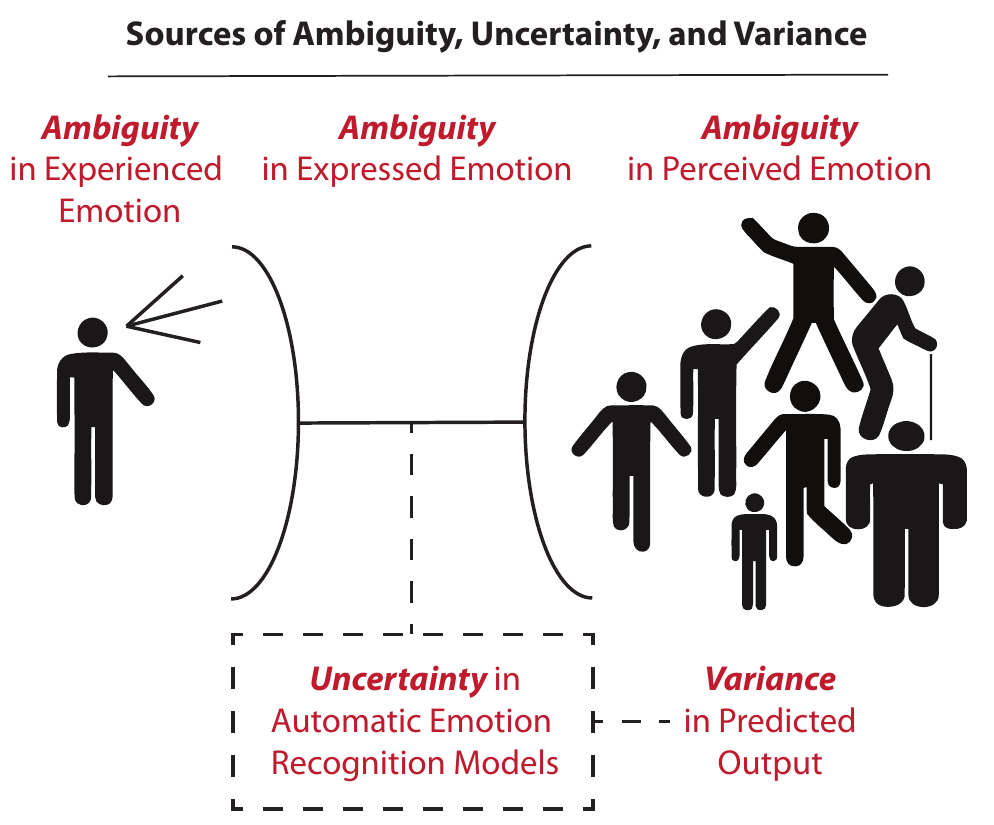}
  \end{center}
 \tiny \caption {An illustration of the Brunswik functional lens model depicting the expression and perception of emotions and the distinction between ambiguity and uncertainty, both of which lead to variance in the output of automatic emotion recognition systems.
 }\label{fig:bigpicture}
  
\end{figure}

\section{Description and Limitations of Existing Emotion Representations}
\label{s:rep_emo}

Affective representations must be grounded in psychological theories of emotion expression and perception in order to circumvent the limitations discussed thus far.  This section provides an overview of the common psychological theories of emotion (Section~\ref{sec:emo_theory}) and standard methods of implementing these theories (Section~\ref{sec:emo_practice}).

\subsection{Theories of Emotion}
\label{sec:emo_theory}
There is no universally accepted theory of emotions.  Several different theoretical accounts of emotions exist, although often fragmented and focusing on distinct aspects (e.g., biological, psychological, philosophical). 

The categorical label representation for emotional state, popular in emotion recognition, has some foundation in theory going back to the Darwinian perspective of universality of basic emotions (such as the six -- happiness, sadness, anger, fear, surprise, and disgust -- identified by Ekman as realised in facial expressions). Various emotion theory researchers have proposed lists of emotion categories (i.e., linguistic labels), typically with under 20, offering plausible representations for computing; a summary of the leading proposals can be found in Cowie and Cornelius \cite{Cowie_2003}. Circumplex models offer a way of capturing conceptual similarity between emotion categories, represented in a circular visual representation with proximally situated categories being similar to one another \cite{plutchik1980emotion, russell1980circumplex}.

The dimensional representation of emotional states, which also has a rich history in the psychology literature, is especially suitable for continuous graded representations, especially for human ratings. An early account by Schlosberg \cite{Schlosberg1941, Schlosberg1954} considered graded representations of judgments of facial expressions and led to the proposal of a two-dimensional axis (pleasantness and attention). The later proposal of Mehrabian and Russell \cite{mehrabian1974} offers a three-dimensional representation -- pleasure, arousal and dominance. These dimensions have been based on factor analysis of experimental data.  Other alternative proposals include those from Scherer \cite{Scherer1984}, along the two dimensions of positive/negative evaluation and activity, and the positive affect and negative affect dimensions of Watson and Tellegen \cite{watson1985}. 

The classic review by Cowie et al. \cite{cowie2001}, especially geared toward computational approaches, summarises a two-dimensional, circular space. One of its axes corresponds to the evaluation (from negative to positive) dimension, interpreted from within cognitive emotion theory as a simplified description of appraisal \cite{Scherer1984} with valence of stimuli considered fundamental to all appraisals \cite{ortony1990}. The other axis is the activation (from passive to active) dimension, with theoretical inspiration drawn from the action tendency theory \cite{frijda1986}. It should be noted that the names given to these dimensions are somewhat varied across the different theoretical accounts, with terms activation, arousal, and activity used interchangeably; and similarly valence, pleasure, and evaluation; and dominance and control dimension. In this paper we will use the terms \textit{arousal} and \textit{valence}. Such multi-dimensional descriptions have been shown to benefit from a greater level of generality across a range of studies, allowing for describing the intensity of emotions including changes over time. These properties are necessary for an analysis of variability in emotion expression for both recognition \cite{grimm2007primitives} and synthesis \cite{schroder2001acoustic}.

Another important aspect lies in relating these (internal latent) emotional state representations to observable signals in human communication and interaction. An adapted version of the Brunswick functional lens model  \cite{Scherer_2003} offers a conceptual framework for representing the expression and perception in interpersonal interaction. Within this framework, the behaviour (e.g., vocal, visual) of a target speaker forms the basis of perceivers' judgements about the target's true latent state (e.g., emotion). This model provides a theoretical framework that describes the expression, communication, and perception process. In this model, there are at least two participants: the speaker (e.g., left-most in Figure \ref{fig:bigpicture}) and the perceiver(s) (e.g., the right-most in Figure \ref{fig:bigpicture}). The speaker produces and encodes an emotional message as distal indicators. The distal indicators may be either unimodal (e.g., the voice) or multimodal (e.g., the face, voice, body gestures, etc.) and are then transmitted to a perceiver. The transmitted information is referred to as proximal percepts. The proximal percepts are decoded by the perceiver and the perceiver makes a perceptual judgment.

In summary, models of human annotation of perceived emotions can target emotion state representations, and access them through their observable expressions guided by a rich history of established emotion theories.

\subsection{Computational Representations of Emotion}
\label{sec:emo_practice}

From the perspective of automatic emotion recognition, the focus has been largely on seeking representations of emotions (especially in defining constructs that are the targets for computation) that can be derived from observations of emotional expressions. These have been guided by theory but often motivated by the ultimate practical application of the recognition technology (e.g., in detecting frustration), especially given there are no broadly agreed upon set of representations, and in how they are behaviourally realised across social-cultural contexts. One of the key challenges in defining a theoretically-grounded construct for computation is the wide range of variation in emotion episodes, from fully blown emotions to varying shades of emotional colouring. The term ``emotional state'' (and emotion-related state) proposed by Cowie and Cornelius \cite{Cowie_2003} is especially apt for modelling emotions. The question then relates to what are plausible emotional representations. 

Emotion representations are inherently tied to the time scale that they describe.  Common time-scales include conversation-level, speaker turn-level, and sentence-/utterance-level.  Studies have used shorter lexical units to increase the temporal resolution (e.g., FAU AIBO corpus was annotated at the word level \cite{Batliner_2008}). Dimensional labels are annotated either over discrete segments or in a time-continuous manner.  Categorical labels are often annotated over discrete segments of speech.  One exception is the SEMAINE database, which used time-continuous traces for categories such as anger, disgust, happiness and amusement \cite{McKeown_2012}. However, these categories were not consistently annotated for all recordings. 

In the remainder of this section, we discuss the collection paradigms for the common emotion representations and their underlying challenges.

\subsubsection{Categorical Labels}

Categorical labels are commonly collected using multiple choice surveys. This provides a discrete set of labels.  

One of the challenges associated with categorical labelling strategies is the process of obtaining the labels.  Surveys are deployed with lists of emotion classes.  Yet, the inclusion of certain classes in the survey biases the responses, forcing evaluators to select a specific class \cite{Russell_1993}. Consequently, the class \emph{other} may be included, providing evaluators the opportunity to list other categories. 

A second challenge with categorical labelling is the large number of overlapping emotional states. For example, Cowie and Cornelius \cite{Cowie_2003} listed 38 emotional states. But, these terms are not mutually exclusive \cite{Lotfian_2017} (e.g., happiness and excitement). Studies have addressed this challenge by proposing primary and secondary emotions.  The primary emotion corresponds to the prominent emotional class, and the secondary emotion(s) include all the other affective states perceived in the stimulus \cite{Busso_2017,Lotfian_201x,Devillers_2005,Cowie_2003}. Lotfian and Busso \cite{Lotfian_2018} demonstrated that leveraging secondary emotions can lead to improvements in classification performance in emotion recognition evaluations.

\subsubsection{Numerical Labels}

Valence and arousal are the most popular dimensions used in affective computing. They are commonly collected using Likert-type scales. A common survey includes \emph{self-assessment manikins} (SAMs), non-verbal pictorial representations describing different levels of the emotional attributes. 

There is an inherent relationship between categorical and numerical descriptions of emotion.  We visualise this relationship using sentences from the MSP-Podcast corpus \cite{Lotfian_201x}, labelled with categorical emotions in the arousal-valence space (the corpus was annotated with numerical and categorical labels). The figure shows that sentences labelled with the same emotional category often span broad areas in the arousal-valence space.  Using emotional attributes provides a powerful representation to quantify the within-class variability of categorical emotions (e.g., different shapes of happiness). 

One of the challenges associated with numerical labelling is most salient given time-continuous annotations.  In this case, evaluators perceive the emotional content in real-time and respond by moving the cursor of the \emph{graphical user interface} (GUI) to the perceived value. Examples of toolkits for time-continuous evaluators are Feeltrace \cite{Cowie_2000}, Gtrace \cite{Cowie_2013}, and CARMA \cite{Girard_2014}. The advantage of this approach is the higher temporal resolution of the emotional annotations, which provides local information to identify emotional segments. A disadvantage of this approach is the reaction lag between the emotional content and the annotations provided by the evaluators, which can be higher than three seconds \cite{Nicolle_2012}. Fortunately, there are algorithmic solutions to compensate for this reaction lag using both static delays \cite{Mariooryad_2013_3, Mariooryad_2015} and delays that vary as a function of the acoustic space \cite{khorram2019lag}. 

\begin{figure}
 \begin{center}
    \includegraphics[width=0.5\textwidth]{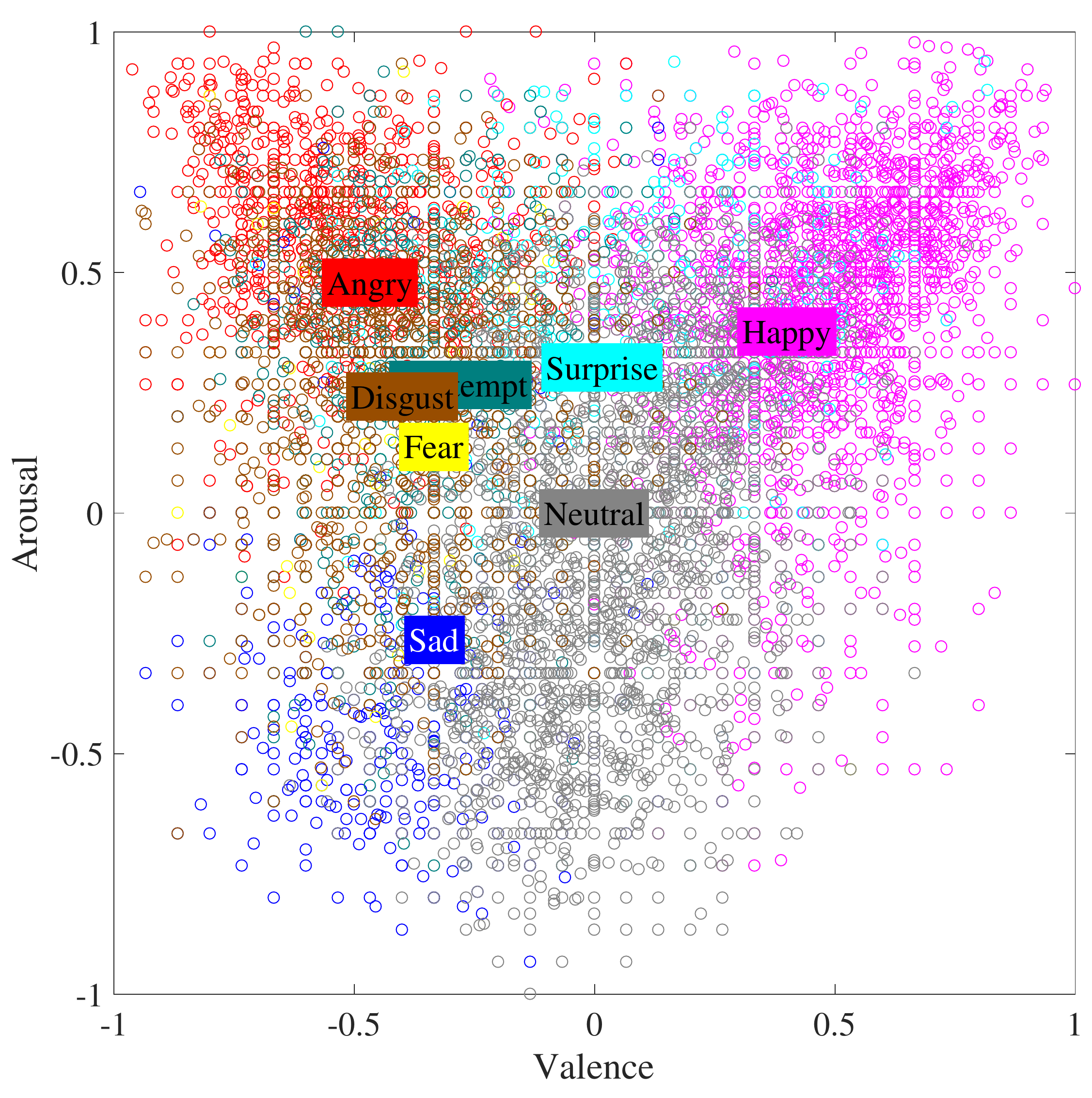}
  \end{center}
 \caption {Categorical labels of sentences from the MSP-Podcast corpus on the arousal-valence space. The categorical labels are placed at the centroids of their sentences. The figure illustrates the relationship between categorical and numerical labels.} 
 \label{fig:arousalvalence}
\end{figure}

A second challenge is that valence and activation alone are not enough to fully characterise all emotional behaviours. For example, different emotional categories such as \emph{anger} and \emph{fear} present similar values for arousal and valence (i.e., low valence, high arousal). Figure \ref{fig:arousalvalence} also illustrates this observation. However, many affective computing datasets contain very small amounts of these overlapped emotions (e.g., fear), which has lessened the computational focus on this problem. 

\subsubsection{Ordinal Labels}
\label{s:rep_ord}

Ordinal labels have a well-defined order but no notion of distance between the labels. They result from direct comparisons between two or more stimuli. This approach can be implemented with either attributes (e.g., is video one more positive than video two?), or with categorical classes (e.g., is video one happier than video two?). Several studies have argued that comparing two or more stimuli is more reliable than separately assigning absolute scores or labels to a given stimulus. Yannakakis et al. \cite{Yannakakis_201x,Yannakakis_2017} presented arguments supporting the ordinal nature of emotions, providing clear evidence across domain within affective computing. Their conclusion was that ordinal labels increase the validity and reliability of affective labels, leading to better models.

Ordinal labels are often derived from categorical or numerical labels. However, they can also be directly obtained with perceptual evaluations. AffectRank \cite{Yannakakis_2017} is an example of a toolkit to directly annotate ordinal labels. Ordinal labels can be directly used to train preference learning models such as RankNet \cite{Burges_2005} and RankSVM \cite{Joachims_2006}.

\subsection{Limitations}
As outlined in this section, a number of schemes for representing emotions have been proposed, each with a different set of pros and cons. The lack of consensus on a single scheme and the sheer number of various schemes that have been employed reflect the lack of a settled theory of emotions, the complexity of emotions, and the emphasis of differing properties of emotions in different studies. A major stumbling block to efforts to converge to fewer but more powerful representations is a lack of a common framework within which current schemes can all be compared and their implicit assumptions in each of them examined and compared.

Additionally, as mentioned in section \ref{s:intro}, most current work in affective computing employs emotion representation schemes that circumvent any ambiguity in perceived emotions. Even the few studies of ambiguity-aware emotion representations suffer from the lack of well motivated methods for measuring accuracy and comparing with each other. This is once again due to the lack of a common description language.

\section{Proposed AMBER Framework}

As outlined in section \ref{s:rep_emo}, a number of approaches have been employed to represent emotions. Broadly they are all descriptions of emotions defined on one of three possible spaces: (a) space of categorical emotion labels; (b) space of numerical affect attributes; or (c) space of ordinal affect labels. In this section we develop the \emph{AMBiguous Emotion Representation} (AMBER) framework, a unified mathematical description within which the myriad of emotion representations defined, in any of these three spaces of emotion labels, can be described. The AMBER framework can be thought of as broadly comprising of two components: (a) a common representation scheme based on a set of suitable descriptors, which we refer to as \emph{attribute descriptors}, that can describe all three possible spaces of emotion labels (\emph{categorical}, \emph{numerical} and \emph{ordinal}); and (b) a function based representation of emotion defined on the attribute descriptors to encode ambiguity.

\subsection{Representation based on Attribute Descriptors}
The benefits of a single framework that can mathematically describe all three kinds of emotions labels include the ability to: (1) reason about known emotion representation schemes; (2) compare them; and (3) identify implicit assumptions made when employing them. We begin with the observation that all three types of emotion labels can be described using a finite number of possible labels (e.g., one of \textit{big six} categorical labels) or a finite dimensional vector (e.g., a 2-dimensional vector comprising of an \textit{arousal} and a \textit{valence} scores). Based on this, we build a common representation framework on the notion of \emph{attribute descriptors}, a finite number of which will be used to describe any space of emotion labels. 

Mathematically, each attribute descriptor is simply a set on which we impose additional mathematical constraints to reflect the properties of the emotion label space of interest. Specifically, if we require $N$ attribute descriptors to describe an emotion label space, then each attribute descriptor, $X^{(n)}$, is defined as an ordered set of elements such that,

\begin{equation}
X^{(n)} = \left\{x^{(n)} \, \middle | \, \alpha^{(n)} \preceq x^{(n)} \preceq \beta^{(n)} \right\}, \qquad 1 \leq n \leq N, \label{e:set}
\end{equation}
where $n$ denotes that $X^{(n)}$ is the $n^{th}$ attribute descriptor;  $\alpha^{(n)}$ is the lowest element in the set $X^{(n)}$ (precedes all other elements); and $\beta^{(n)}$ is highest element in the set $X^{(n)}$ (is preceded by all other elements).

To illustrate how emotion attribute descriptors represent familiar quantities in common emotion representation schemes, consider two examples.  First, we consider an example where emotional state is represented as a point on an \textit{arousal-valence} plane.  The first \emph{affect attribute}, $X^{(1)}$, could denote \textit{valence} and the second one, $X^{(2)}$, could denote \textit{arousal}.  If we assume that both valence and arousal vary between -1 and 1, then the lowest element in the set ($\alpha^{(n=1,2)}$) would typically be $-1$ and the highest element in them ($\beta^{(n=1,2)}$) would be $1$.  All other elements of the set are the real numbers between $-1$ and $1$. i.e., $x^{(n=1,2)} \in [-1,1]$. 

Second, we consider an example where emotional state is represented as one of a set of emotional categorical labels, such as the \textit{big six}.  Here, we could employ six \emph{affect attributes} (i.e., $N=6$) and each one, $X^{(n)}$, would denote one of the six possible categorical labels, \textit{happy}, \textit{surprised}, \textit{afraid}, \textit{sad}, \textit{angry} and \textit{disgusted}. Each \emph{attribute descriptor} would then be a two element set, $X^{(n=1,\ldots,6)}=\left\{ \mathcal{O}, \mathcal{I} \right\}$ such that $\mathcal{O}$ and $\mathcal{I}$ denotes the absence and presence of a corresponding categorical label. A typical categorical label would then be a sparse set of attribute descriptors with only one of them indicating the presence of a categorical label (non-binary variations corresponding to blended emotion labels and mixtures are introduced in section \ref{sec:AMBER_ambiguity} and examples are discussed in section \ref{sec:cat_labels}).

In general, given a set of $N$ affect attributes, the emotion at time $t$ is represented by a set of $N$ elements, $\Psi(t)$, as follows:

\begin{equation}
\Psi(t) = \left\{ \tilde{x}^{(n)}(t) \, \middle| \, n = 1, \ldots , \, N \right\}, \label{e:rep}
\end{equation}
where each $\tilde{x}^{(n)} \in X^{(n)}$. 


AMBER permits an explicit statement of the mathematical structure underlying emotion representations. This is important, because these statements, and the associated assumptions, are more often implicit.  Consider a common treatment of the arousal-valence space (Figure \ref{fig:arousalvalence}): clustering via Euclidean distance.  This treatment makes two assumptions: (1) that the appropriate metric is Euclidean and (2) that the two axes are orthogonal.  These assumptions are not inherently incorrect, but they must be stated.  The AMBER framework provides a language to do so.  For example, the affect attributes $X^{(1)}$ and $X^{(2)}$, denoting \textit{valence} and \textit{arousal}, are both mapped to $[-1,1]$ on the real number line.  Second, they are orthogonal to each other and the standard Euclidean distance metric exists in the space $X^{(1)} \times X^{(2)}$.

\subsection{Ambiguity Awareness in the Proposed Framework}
\label{sec:AMBER_ambiguity}
The representation given by equation (\ref{e:rep}) can describe emotion representation schemes employing any of the three types of emotion label spaces.  However, it is restricted to single-valued representations only. This restriction can be overcome by removing the constraint that the representation should be given by only one element chosen from each \emph{attribute descriptor set}, $X^{(n)}$. Specifically, we can extend it by allowing the representation to be a set of functions, one per attribute descriptor that denotes a combination of multiple elements of that attribute descriptor set.

Let us consider the emotion profile representation \cite{mower2011framework} as an example to illustrate the use of a suitable ambiguity aware representation. This scheme represents the emotion in terms of a finite set of categorical labels (\emph{Angry}, \emph{Happy}, \emph{Neutral} and \emph{Sad}) but instead of picking one of them, the emotion is represented as the set of probabilities of each one being the right label. When this scheme is described using the proposed framework, the emotion label is a set of four probabilistic measures (on for each attribute descriptor) that indicates the probability of presence of that emotion category, $\Psi = \left\{P\left(x^{(n)}=\mathcal{I}\right) \middle| n = 1,\dots,4\right\}$. Where, each attribute descriptor, $X^{(n=1,\ldots,4)} = \left\{ \mathcal{O}, \mathcal{I} \right\}$, corresponds to one of the four categorical labels. Each of these sets comprises of two elements denoting the presence or absence of that emotion category.

Mathematically, this proposed extension, which allows ambiguity in the representations to be quantified, defines the representation of the emotional state, $\Psi(t)$, as
\begin{equation}
\Psi(t) = \left\{ \xi_{n,t} \left(x^{(n)}\right) \, \middle| \, n = 1, \ldots , \, N \right\}, \label{e:unc_rep}
\end{equation}
where $\xi_{n,t}: X^{(n)} \rightarrow [0,\infty)$ denotes a function that associates each element of $X^{(n)}$ with a positive real number at time $t$.  We can compare two labels, $x_i^{(n)}$ and $x_j^{(n)}$ at time $t$ using this function.  If $\xi_{n,t} \left( x_i^{(n)} \right) > \xi_{n,t} \left( x_j^{(n)} \right)$, for $x_i^{(n)}, x_j^{(n)} \in X^{(n)}$, we can say that it is more likely that $x_i^{(n)}$ represents the emotion at time $t$ compared to $x_j^{(n)}$. This function is herein referred to as the \emph{ambiguity function}. Examples of emotion representation schemes that take into account ambiguity in labels and how they may be viewed in terms of the framework introduced here are discussed in section \ref{s:AMBER_rep}.

This extension in (\ref{e:unc_rep}) generalises equation (\ref{e:rep}) by relaxing the assumption that a single label per attribute descriptor adequately represents perceived emotion by assigning a level of certainty to every possible label. Specifically, the emotion representation given by equation (\ref{e:rep}) can be seen as equivalent to a special case of that given by equation (\ref{e:unc_rep}) where,
\begin{equation}
\xi_{n,\cdot} (x) =
\begin{cases}
1, & x=\tilde{x}^{(n)} \\
0, & x \neq  \tilde{x}^{(n)}, \label{e:certainty}
\end{cases}
\end{equation}
i.e., emotion representations of the form indicated by equation (\ref{e:rep}) make the assumption that there is no ambiguity about the perceived emotional state, which is clearly known to be incorrect as discussed in section \ref{s:intro}. An illustration of this distinction in the context of categorical labels is shown in Figure \ref{fig:cat_amber}.

\section{Reinterpreting Common Representations through an AMBER lens}
\label{s:AMBER_rep}

As mentioned in section \ref{s:rep_emo}, a number of methods have been employed in the literature to represent emotions. This section will illustrate how all of these methods can be reinterpreted in terms of the proposed AMBER framework. This in turn will allow the explicit as well as implicit assumptions behind these methods to be viewed and compared within a common mathematical framework and, consequently, allow us to assess their suitability to various contexts.

\subsection{Categorical Labels}
\label{sec:cat_labels}
We will begin by evaluating the categorical labelling scheme, where a single label is assigned to an interval of interest, through the lens of AMBER. First, each emotion dimension, $n$, denotes a different emotion category (for e.g., $X^{(1)}$ denotes \textit{Anger}, $X^{(2)}$ denotes \textit{Happy}, etc.) with each emotion dimension given by a two element set,

\begin{equation}
X^{(n)} = \left\{ \mathcal{O}, \mathcal{I} \right\}, \qquad 1\leq n \leq N,
\end{equation}
where $\mathcal{O}$ and $\mathcal{I}$ denote the absence and presence of the $n^{th}$ emotion category.

Second, the emotion labels are not time-varying within the intervals of interest. For example, when an utterance is labelled as \textit{Happy} it is generally assumed that this emotion category applies at all times within the entire interval of interest, i.e.,
\begin{equation}
\xi_{n,t} = \xi_n , \qquad \forall  t.
\end{equation}

Finally, in many cases there is an implicit assumptions that the categorical emotion labels are chosen from a finite set of possible emotion dimensions (categories) and these are all mutually exclusive (see \textit{Box 1} for analysis of a representation scheme where this assumption is relaxed), i.e.,
\begin{equation}
\Psi = \left\{ \tilde{x}^{(n)} \middle| \forall n; \ s.t. \ \tilde{x}^{(n)}=\mathcal{I} \ \text{for only one n} \right\}
\label{e:cat_certain}
\end{equation}

It is clear when viewed through the AMBER framework that all three assumptions impose well-defined mathematical restrictions on the emotion representation scheme.  But, the framework also suggests how these assumptions may be relaxed to obtain more generalised representations. For instance, relaxing the constraint given by equation (\ref{e:cat_certain}) that the emotion dimensions are mutually exclusive and allowing more than one dimension to take on the value $\mathcal{I}$ opens up the path for categorical emotion labelling schemes where labels can comprise multiple emotion categories for a single particular time interval.

\begin{mdframed}
\subsubsection*{Box 1: Blended Emotions}
The notion of `blended emotions' as described by Vidrascu and Devillers is one such method where emotions are represented using both a `major' and a `minor' categorical label \cite{vidrascu2005real}. Here, up to two emotion dimensions may take on the value $\mathcal{I}$ and some ambiguity is allowed. Specifically, it can be seen that the `blended emotions' representation may be given as

\begin{equation}
    \Psi_{blended}=\left\{ \xi_n\left(x^{(n)}\right) \middle| n=1, \ldots , N\right\}
\end{equation}
with three additional constraints. Firstly,
\begin{equation}
    \sum_{n=1}^{N} \xi_n(\mathcal{I}) = 1
\end{equation}
where the assumption $\xi_n(\mathcal{O}) = 1 - \xi_n(\mathcal{I}), \ \forall n$ allows for $\xi_n$ to be treated as a probability distribution without any loss of generality. Secondly,
\begin{equation}
    \sum_{n=1}^{N} \left[\xi_n(\mathcal{I})>0 \right] = 2
    \label{e:two_cat_constr}
\end{equation}
where $[\cdot]$ denotes the Iverson bracket. i.e., $[condition]$ takes the value $1$ if $condition$ is true and the value $0$ if $condition$ is false. This condition constrains the representation to only two labels at most. Finally, $\xi_n(\mathcal{I})$ takes on one of two values, i.e., $\xi_n(\mathcal{I})\in\{p,q\}$ with $p<q$ and
\begin{equation}
    \xi_n(\mathcal{I}) = 
    \begin{cases}
    p, & minor\ emotion \\
    q, & major\ emotion
    \end{cases}
\end{equation}
This conditions ensures that one of the two labels denotes the major  emotion and the other the minor emotion.
\end{mdframed}

The emotion profile representation \cite{mower2011framework} is an more general categorical emotion representation which makes fewer implicit assumptions, as can be seen when viewed within the AMBER framework. Specifically, the emotion profile representation drops the constraint given by equation (\ref{e:two_cat_constr}). i.e., it represents emotions as a set of ambiguity measures, each corresponding to a different emotion categorical label.

\begin{figure*}
 \begin{center}
      \includegraphics[width=0.7\textwidth]{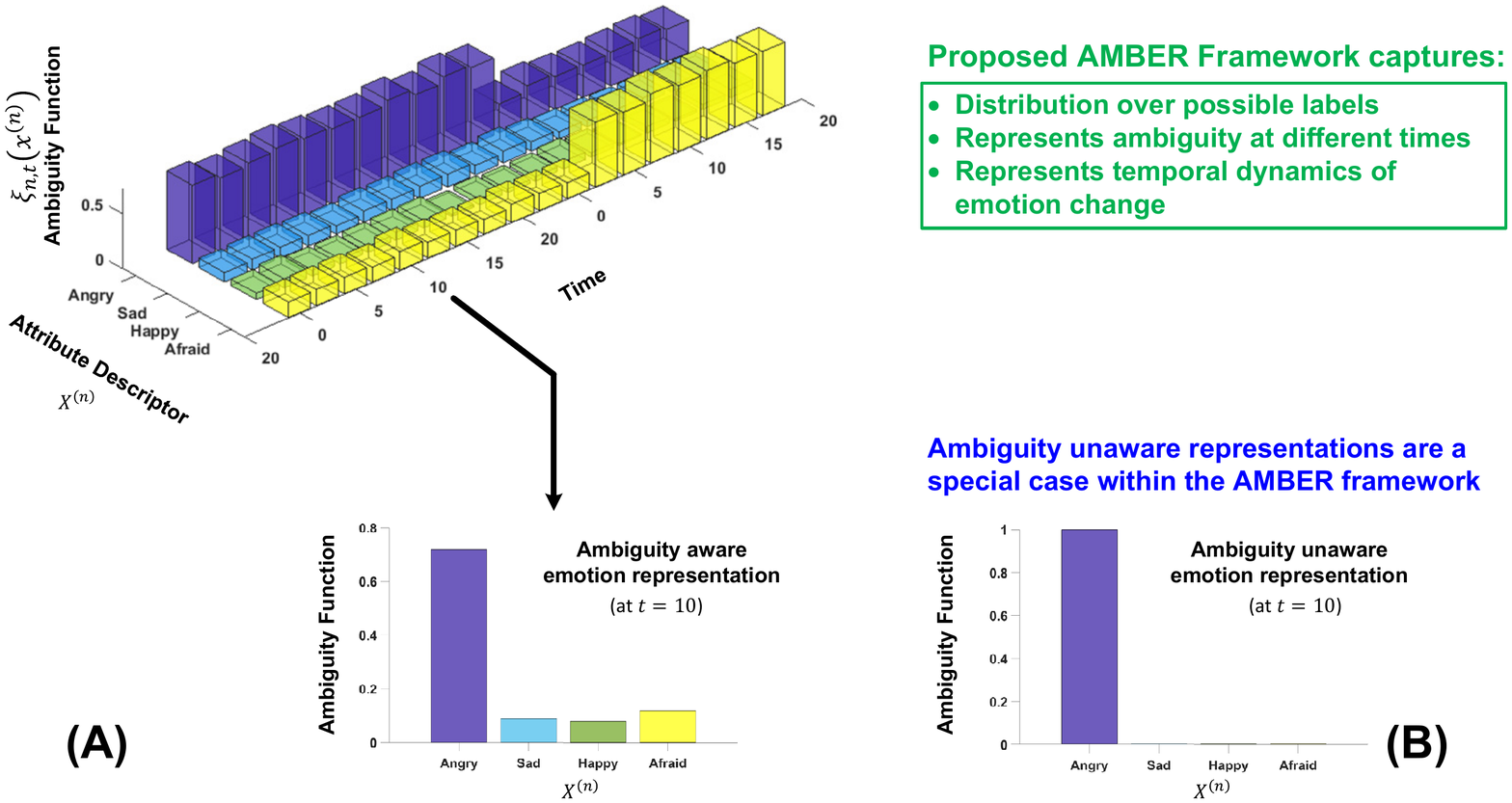}
  \end{center}
 \tiny \caption {A depiction of the differences between ambiguous and non-ambiguous categorical emotion representations within the proposed AMBER framework: (A) shows an ambiguity aware label at a specific time ($t=10$) given by the set of probabilities of the presence of each of the four possible emotion category labels; (B) shows that assuming a single categorical label is akin to complete certaintly}\label{fig:cat_amber}
\end{figure*}

The most general emotion representation that can be defined within a categorical label space within the AMBER framework is depicted in Figure \ref{fig:cat_amber}. This is essentially a set of measures of ambiguity of a finite number of emotion categories that varies with time.

\subsection{Numerical Attribute Descriptors}
Numerical attribute descriptors are probably the most widely used emotion representation schemes today in the field of affective computing. They can be described within the proposed framework by introducing a notion of distance, $d_{i,j}^{(n)}$, between pairs of elements within each attribute descriptor set, $X^{(n)}$, each of which represents one of the numerical attributes of interest. Specifically, the bounded ordered set $X^{(n)}$ is mapped to a closed interval $[a,b]$ on the real number line via a suitable function $f_n:X^{(n)}\rightarrow[a,b]\in \mathbb{R}$. After mapping the emotion into a number, we can adopt the natural metric on real numbers as the distance between emotions ($d_{i,j}^{(n)}$), i.e.,

\begin{equation}
d_{i,j}^{(n)} = \left| f_n\left(x_j^{(n)}\right) - f_n\left(x_i^{(n)}\right) \right| \label{e:distance}
\end{equation}

This converts the ordered set $X^{(n)}$ (denoting \textit{arousal}, \textit{valence}, \textit{dominance}, etc.) to a numerical scale and commonly the interval $[a,b]$ is chosen as $[-1,1]$. Disagreements concerning the representation of emotional state as perceived by different annotators can arise from differences in the perceived emotional state (and/or from differences in their mapping of emotional state to numerical attribute score - refer Box 2). This disagreement in turn may reflect important information such as the level of difficulty in perceiving the emotional state, and condensing them to a single-valued emotion representation (i.e., when equation (\ref{e:certainty}) of the AMBER framework holds) does not capture this information.

\begin{mdframed}
\subsubsection*{Box 2: Annotator Transcription Normalisation}
Some of the implications of adopting a numerical attribute emotion representation scheme are now much more explicit when viewed within the AMBER framework. Namely, this scheme typically assumes that different annotators transcribe the perceived emotion attributes identically, i.e., $f_n$ is identical for all annotators. If this assumption does not hold (and there is no obvious reason why it should), then it might be prudent to normalise for differences between annotators in terms of how they transcribe. Hypothetically, such a normalisation  may be viewed as an attempt to learn annotator specific maps, $f_{n,p}: X^{(n)} \rightarrow [a,b]$, corresponding to the $p^{th}$ annotator, and modify them to obtain normalised labels, $\hat{X}^{(n)}=f_n^{-1}\left( f_{n,p}\left( X^{(n)} \right) \right)$. The authors are not aware of any work to date that explicitly attempts this normalisation. However, an interesting approach to tackle the differences between annotators was reported by Grimm and Kroschel, whereby confidence in each annotator was estimated and the confidence measures were employed as weights when estimating a single-valued label as a weighted sum of the individual labels \cite{grimm2005evaluation}.

\end{mdframed}

\begin{figure*}
 \begin{center}
      \includegraphics[width=0.9\textwidth]{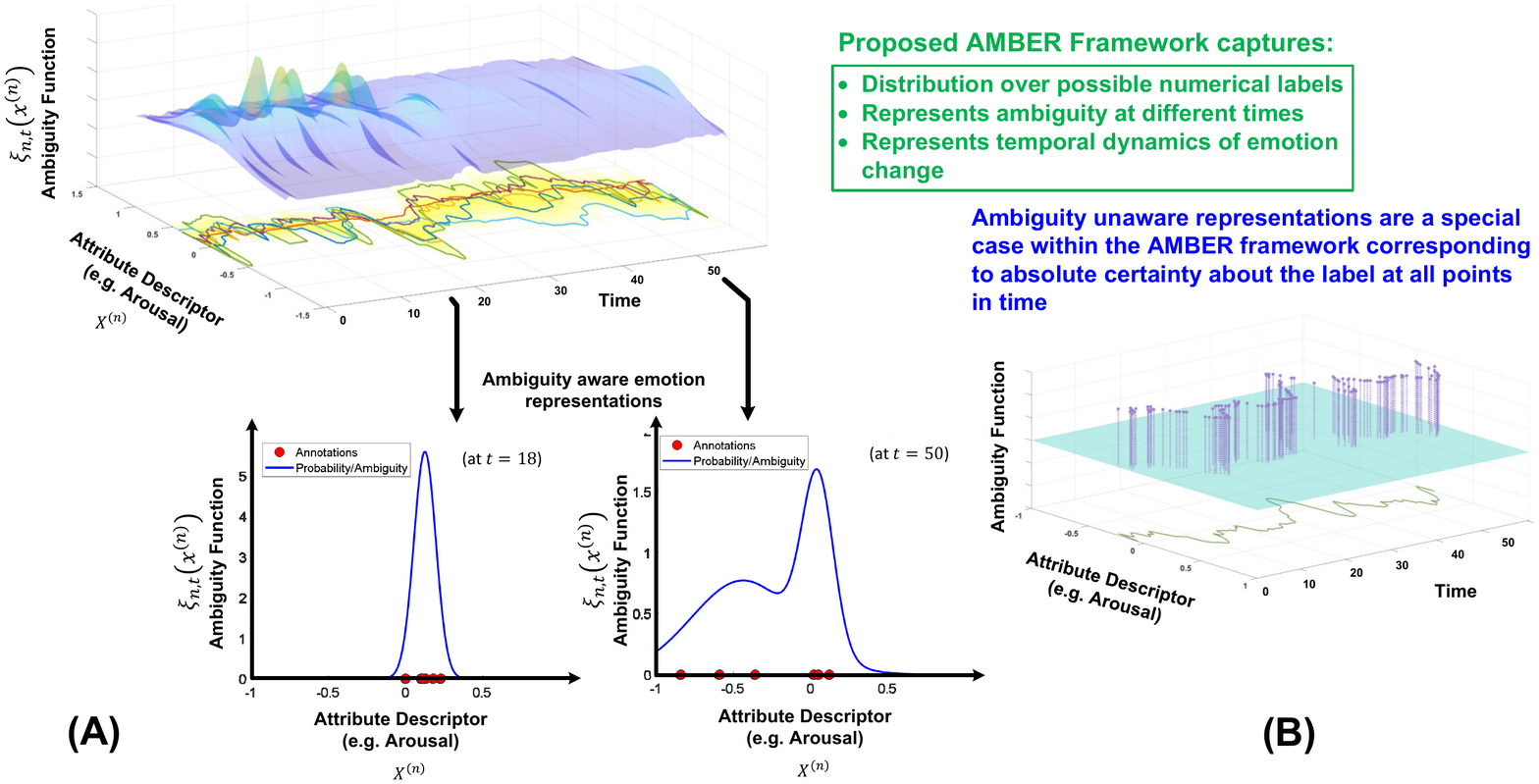}
  \end{center}
 \tiny \caption {(A) A time-varying numerical emotion representations within the proposed AMBER framework based on numerical annotations obtained from multiple annotators (individual traces on the Arousal-Time plane) is given by a time varying distribution (ambiguity function depicted in 3D as a function of time, each slice is a distribution at a particular point in time); (B) an ambiguity unaware representation, such as using the mean annotation across all raters, ignores time-varying disagreement between annotators and the corresponding ambiguity functions are dirac deltas reflecting this.}
 \label{fig:num_amber}
\end{figure*}

The time-varying numerical attribute labels also employ the distance measure given by equation (\ref{e:distance}) but the emotion representation $\Psi(t)$ additionally retains their dependence on time.  Figure \ref{fig:num_amber} shows an example of ambiguity aware time-varying arousal labels, wherein the ambiguity function at each time $t$ captures the ambiguity inherent in the set of arousal traces obtained multiple annotators. Furthermore, the abiguity function can be any distribution, including multimodal ones. This figure corresponds to the emotion labelling scheme proposed in \cite{dang2017investigation}.

In addition to the implications arising from adopting a distance metric within each attribute descriptor set, a few additional implications arise from the dependence on time. Namely, the distance metrics $d_{i,j}^{(n)}$ are almost universally assumed to be time-invariant or equivalently $f_n$ does not change with time. However experience tells us that annotators may not give identical numerical attribute scores to identical stimuli on two separate occasions. It is also common to not explicitly take into consideration constraints on the temporal dynamics of the labels, i.e., conditions on the relationship between $\Psi(t)$ and $\Psi(t-\epsilon)$ for any $\epsilon$. Clearly, constraints can be explicitly considered, since emotional state will not change instantaneously, but the specific forms these constraints should take is still an open research question.

\begin{mdframed}
\subsubsection*{Box 3: Numerical Attribute Label Distributions}
In recent years there has been growing interest in accounting for the ambiguity in emotion labels reflected in the disagreement between annotations. While a number of approaches to handle this have been proposed, they all essentially reduce to the emotion representation given by equation (\ref{e:unc_rep}) with some additional constraints. For instance, the scheme employed by Han et al. \cite{han2017hard} assumes that the ambiguity function, $\xi_n(\cdot)$, is a Gaussian function and is represented by its mean and standard deviation. A less restrictive assumption is made by Dang et al. \cite{dang2017investigation}, who allow the ambiguity function to be any distribution that can be represented as a Gaussian mixture model. Both approaches were extended to include a model of the temporal dynamics of the ambiguity function, making use of a Long Short-Term Memory Recurrent Neural Network \cite{han2017hard} and a Kalman filter \cite{dang2018dynamic} respectively. Finally, Atcheson et al. \cite{atcheson2018demonstrating, atcheson2017gaussian} strongly suggest that time-varying annotations should be treated as stochastic processes.
\end{mdframed}


\subsection{Ordinal Labels}
The defining feature underlying all ordinal labelling schemes is that there is a notion of order between any pair of labels but there is no notion of distance between them. That is, for any pair of elements within an emotion dimension, $x^{(n)}_i,x^{(n)}_j \in X^{(n)}$, one can say whether $x^{(n)}_i \preceq x^{(n)}_j$ or whether $x^{(n)}_i \succeq x^{(n)}_j$, but $d^{(n)}_{i,j}$ cannot be defined. Ordinal labelling schemes can be broadly categorised into one of two categories. The first one typically involves labels drawn from a small finite set of ordered elements, such as a Likert-type scale or self-assessment manikins. The second category typically involves labels that reflect a comparison between two or more intervals of interest. This essentially gives rise to a ranking scheme. We refer to the first category as \textit{absolute ordinal labels} and the second as \textit{relative ordinal labels} in this paper.

The \textit{absolute ordinal labels} can be viewed as probably the most general case within the AMBER framework and are given by equations (\ref{e:set}) and (\ref{e:rep}), with typically the only condition imposed on them being that $X^{(n)}$ is a small finite strictly ordered set. The \textit{relative ordinal labels} on the other hand can be viewed as imposing the condition that labels at distinct points in time reflect their rank relative to all other points in time. i.e.,
\begin{equation}
    \tilde{x}^{(n)}(t_i) \in \left\{r_0 \preceq r_i \preceq r_T \right\}, \qquad i = 0, \ldots , T \label{e:ord_rank}
\end{equation}

\noindent where, $t_i$ denotes an interval of interest, $T$ denotes the total number of intervals of interest, and $r_i$ denotes the relative rank of the label within emotion dimension $X^{(n)}$ associated with an interval relative to the labels associated with all other intervals. This condition reflects the observation that if every pairwise ordering relationship of emotion labels at different points in time is known then the set of labels at all points in time can be ranked in order. The labels as a function of time can then reflect this rank order (as indicated by equation (\ref{e:ord_rank})).

\begin{figure}[t]
  \centering
  	\includegraphics[width=0.45\textwidth]{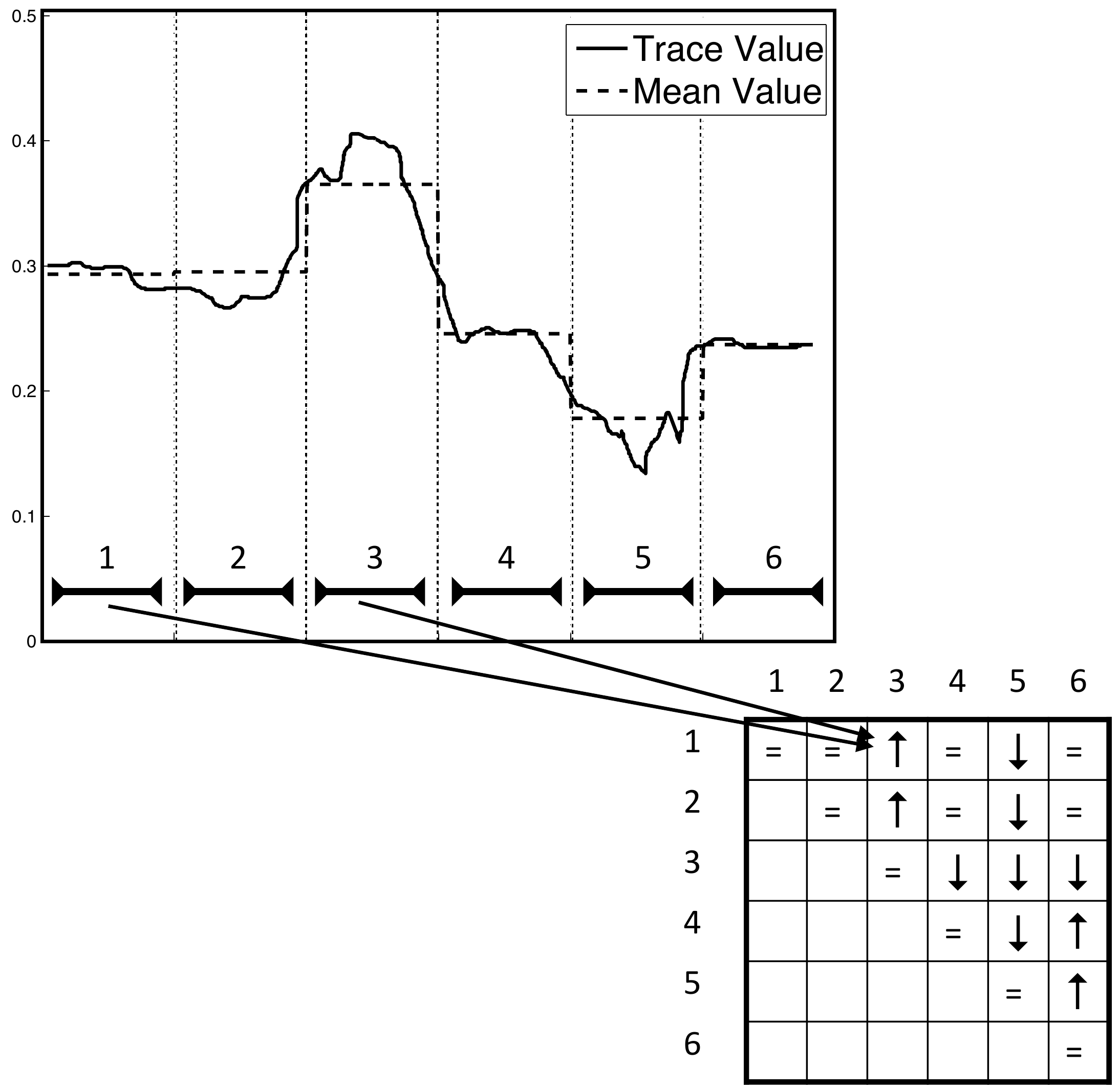}\\
    \includegraphics[width=0.45\textwidth]{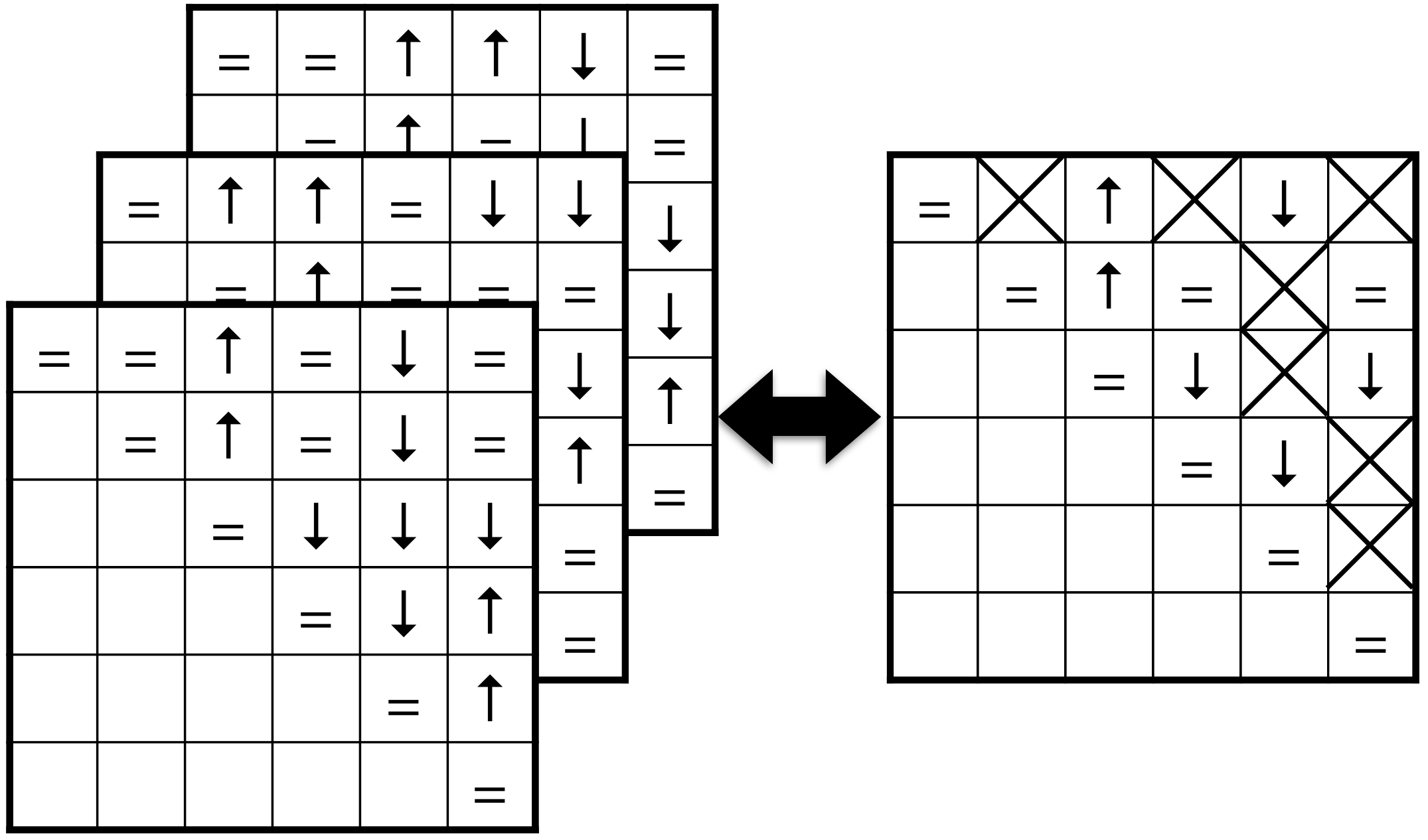}
  \caption{Deriving relative labels from time-continuous traces using the qualitative agreement analysis \cite{Parthasarathy_2016_2}. (A) The individual matrices are created by relative comparison between segments of the trace. (B) The consensus matrix is formed by combining individual matrices. A trend is set when the differences are greater than given thresholds. We cross out the entries without agreement}
  \label{fig:parthasarathy}
\end{figure}

\begin{mdframed}
\subsubsection*{Box 4: Ordinal Labels for Preference Learning}
Recently, Parthasarathy et al. \cite{Parthasarathy_2016_2} demonstrated the benefits of inferring \textit{relative ordinal labels} from time-continuous numerical annotations as targets for a machine learning system. Specifically, they build on the qualitative agreement method where the average numerical label values over intervals of interest are compared pairwise amongst themselves to determine pairwise order relationships. For each trace, the approach generates an \emph{individual matrix} (IM) with the trends in the trace (Figure \ref{fig:parthasarathy}A). Then, the consensus between multiple annotators is ascertained to only retain consistent trends observed across annotators (Figure \ref{fig:parthasarathy}B). The relative labels obtained with this approach lead to a ranked list of all intervals of interest with agreement as given by equation (\ref{e:ord_rank}).
\end{mdframed}

\begin{figure}
 \begin{center}
      \includegraphics[width=0.45\textwidth]{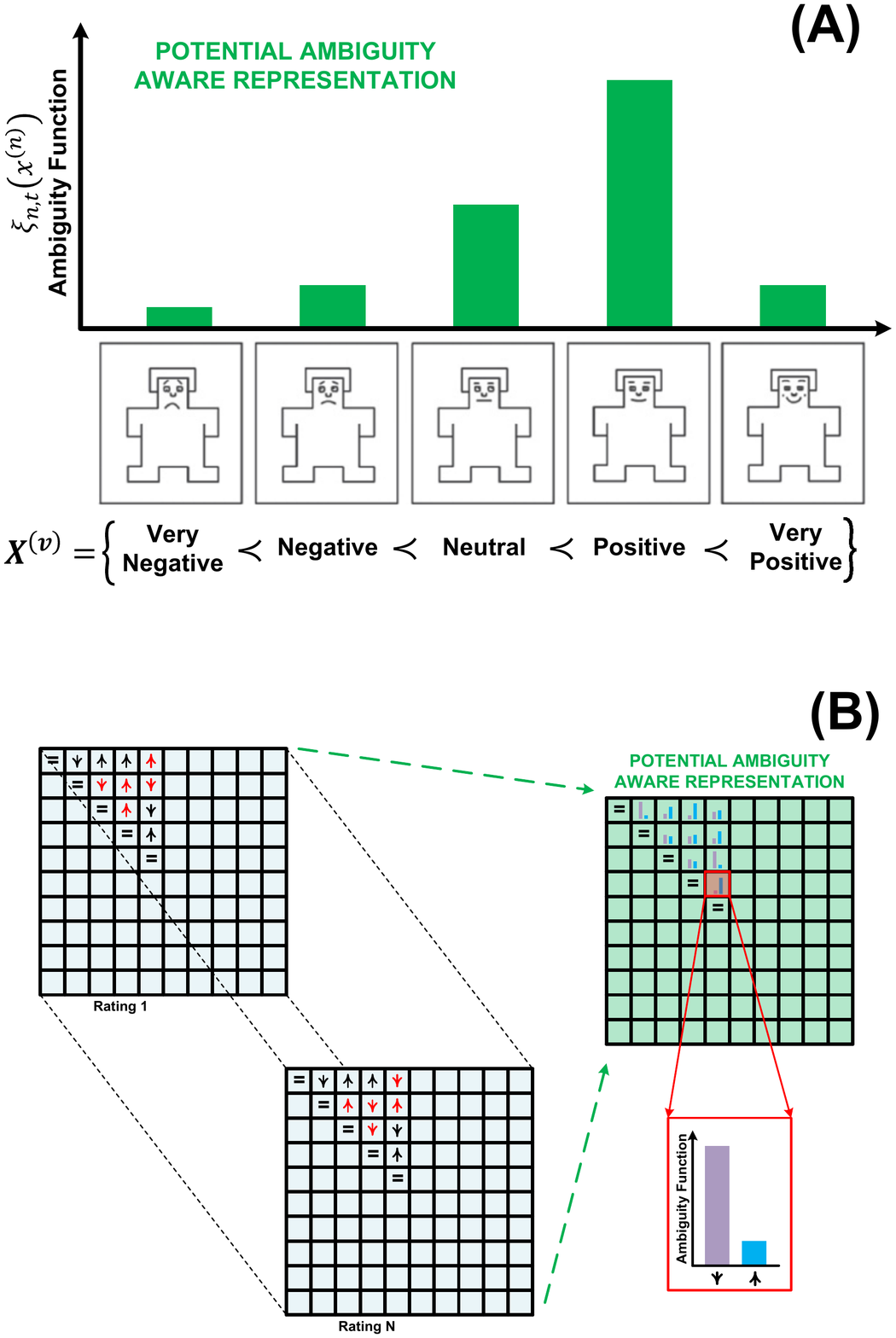}
  \end{center}
 \tiny \caption {(A) Absolute ordinal representation may be made ambiguity aware by defining a distribution over the finite ordered set of possible labels corresponding to each affect attribute. In this figure, the valence attribute, denoted by $X^{(v)}$ is depicted; (B) Relative ordinal labels obtained from pairwise comparisons across time frames may be expanded to be ambiguity aware by defining the ambiguity function as a Bernoulli distribution over the two possible outcomes of each pairwise comparison.}
 \label{fig:ord_amber}
\end{figure}

As previously mentioned in section \ref{s:rep_ord}, it has been suggested that ordinal labels exhibit greater validity and reliability compared with numerical and categorical ones. This has been attributed to the observation that people can more reliably compare two stimuli than assign an absolute score to a single stimulus. However, this does not mean ambiguity can be completely eliminated by simply adopting an ordinal scale. For instance, as illustrated in \cite{Parthasarathy_2016_2}, when combining individual pair-wise comparison matrices to obtain a consensus matrix, agreement cannot be observed for all entries. A truly ambiguity-aware ordinal representation scheme would also quantify the disagreement at these entries. However, thus far no such scheme has been proposed. Nevertheless, the AMBER framework can still be brought to bear on the absolute and relative ordinal schemes suggesting that by defining suitable ambiguity functions, ambiguity aware ordinal representations may be defined. An illustration of such schemes are shown in Figure \ref{fig:ord_amber}. In the case of \textit{absolute ordinal labels}, ambiguity functions can be defined over attribute descriptor sets in a straightforward manner while keeping in mind that no distance metric can exist in this set. In the case of finite ordered sets, this may simply be a distribution function over the set of elements (as depicted in Figure \ref{fig:ord_amber}A). Defining ambiguity functions over \textit{relative ordinal labels} is not as straightforward but, as discussed in Box 4 (and shown in Figure \ref{fig:parthasarathy}), the labels are obtained from pairwise comparisons across time frames and an ambiguity function may be defined over the possible outcomes of each comparison (see Figure \ref{fig:ord_amber}B), reflecting the relative counts of each outcome accumulated across all the annotators (for example, when comparing the arousal at frame 2 with what at frame 5, if 4 out of 6 annotators state it was higher and the other 2 state it was lower, the ambiguity function could take the value $2/3$ corresponding to higher and $1/3$ corresponding to lower).

\section{Discussion - What purpose can AMBER serve?}
The choice of how emotion is represented plays a key role in affective computing systems. It underpins almost every stage of the design and implementation, ranging from data collection, to the choice of machine learning model, and how the system will be used. In other words, the myriad of choices that needs to be made when developing affective computing systems are all in one way or another affected by implicit and explicit properties of the chosen emotion representation scheme. Providing a suitably precise and flexible mathematical language for describing and reasoning about emotion representation methods is the primary goal of the proposed AMBER framework. AMBER draws on commonalities across the various possible emotion representation schemes and encode them within 3 elements of the framework: (a) The relationship between the different sets of attribute descriptor; (b) the structure of each of the attribute descriptor sets; and (c) the properties of the ambiguity function. By enumerating the mathematical properties of these three elements, one can explicitly identify all the assumptions underlying an emotion representation scheme. This in turn can help better inform data collection paradigms, choice of machine learning models and algorithms, quantitative analyses of emotions and the interpretation of outcomes of affective computing systems. The following examples of hypothetical questions that may arise when dealing with affective computing systems are provided to illustrate the various ways in which the AMBER framework might be useful.

\begin{quote}
    \emph{I am training an continuous time emotion prediction system and my training data has labels from 6 annotators, what loss function should I minimise?}
\end{quote}

This is really a two part question. Presumably the emotion representation scheme (categorical, dimensional, or ordinal) has been already determined, which implies the attribute descriptor sets are known and their mathematical structure is fixed (e.g., if the representation scheme is time continuous values of arousal and valence then there are two attribute descriptors \emph{arousal} and \emph{valence} and both are sets of real numbers that fall in the interval $[-1,1]$). What then needs to be determined is a suitable \emph{ambiguity function} based on the properties of the acquired labels (e.g., how many annotators?, were they independent?, was there any measurement error?). In this case, examples of an ambiguity function that encode the time-varying variability between the 6 annotations include: (a) a Gaussian at each time $t$, with time varying means and standard deviations; (b) a Gaussian mixture model at each time $t$, with time-varying parameters that can also cater for multimodal and non-Gaussian label distributions; (c) a Gaussian process over labels at all times that jointly encode both the distribution of multiple labels and their temporal dynamics (smoothness over time).

Following this, the loss function can be defined as metric that measures the separation between two ambiguity functions, one corresponding to the training labels and one obtained from the prediction system (more than one suitable metric is likely to exist, each with different properties some of which might be more desirable than the other). In this case, if the ambiguity function was chosen as a probability density function, as in the examples mentioned above, Kullback-Liebler divergence might serve as a suitable loss function. Finally, it should be noted that the prediction system may not explicitly predict an ambiguity function but nevertheless an implicit ambiguity function may be inferred from the form of the prediction and the associated loss function. For example, if the system is designed to predict a time-varying mean and variance labels, it might implicitly correspond to the choice of a unimodal symmetric density function (such as a Gaussian) as the ambiguity function.

\begin{quote}
    \emph{I am collecting data to train an emotion prediction system, I can only practically obtain annotations using a Likert type scale for arousal and valence but my application demands a continuous arousal and valence predictions in the range $[-1,1]$. Can I still proceed with my plan? Will something go wrong? How can I reason about this?}
\end{quote}

They key challenge here is that the emotion representation scheme employed to annotate the training data and employed by the system to make its prediction differ in their mathematical structure (which in turn reflects differences in the emotion theory underpinning them). In the both cases, there are two attribute descriptors (\emph{arousal} and \emph{valence}). However in the first case they are both finite sets of ordered elements with no distance metric defined on the elements while in the second case they are both sets of real numbers in the range $[-1,1]$ with the natural distance metric on real numbers defined on them. Consequently, treating the elements of the attribute descriptor sets in the annotation scheme as equidistant points within the attribute descriptor sets in the prediction scheme is not implicitly justified. This does not mean that a mapping from one scheme to another cannot be learned but based on their properties, it is likely that a scheme that exploits the fact that both are ordered sets without explicitly making use of distances within them may be more accurate and will be better justified.

\begin{quote}
    \emph{I have some data that has been annotated with both categorical labels as well as numerical labels using arousal, valence and dominance as dimensions. Can I study the relationship between the categorical labels and clusters in the 3-dimensional arousal-valence-dominance space obtain using k-means to learn about the fundamental relationship between categorical and dimensional labelling schemes?}
\end{quote}

This question essentially translates to one of identifying the map between categorical representation scheme given by a set of binary values attribute descriptors (one for every possible emotion label, with the two elements of each set indicating presence or absence) and the numerical scheme given by a 3 sets of real valued numbers lying in the interval $[-1,1]$. This in itself is a justifiably posed question that is worth asking. However, the adopted method brings with it a number of implicit assumptions, not all of which can necessarily be justified. For instance, clustering the numerical labels in a 3-dimensional space assumes at the very least a suitable distance metric in that space.  Further, typical implementations of the k-means algorithm tend to use a Euclidean distance metric which additionally assumes that the 3 attributes are orthogonal to each other. Finally, the method adopted to learn the relationship between clustered numerical labels and categorical labels might make further assumptions about the attribute descriptor sets (such as the 3-dimensional arousal-valence-dominance space is a vector space and is equipped with notions of scaling and translation). It is worth noting that all of these assumptions may be reasonable ones to make, however, the researcher making them should actively choose to do so and ideally also state them explicitly when reporting studies based on the outcomes of such analyses.

\section{Conclusions}
In this paper, we introduce the AMBER framework to describe emotion representations, including the ambiguity inherent in them, and thereby serving as a mechanism to reason about them. To the best of the authors knowledge, every emotion representation scheme employed in affective computing till date can be described within the proposed AMBER framework. Furthermore, it allows assumptions implicit in these emotion representation schemes to be clearly articulated and in a manner that allows them to be compared across different schemes. Consequently, it provides a means to reason about how to extend representation schemes in order to imbue them with desirable properties. For instance, if one desired an ordinal emotion representation scheme that incorporated ambiguity and would be employed by other AI systems to reason about human-computer interactions within a Bayesian framework, then in terms of the components of the AMBER framework, one would consider equipping a known ordinal scheme with an ambiguity function that was also a probability distribution. The proposed AMBER framework also provides a means to reason about analyses carried out using emotion labels and annotations. For example, is it reasonable to run a clustering algorithm on emotion labels? Should the joint probability distribution over arousal and valence be modelled? Is taking the mean of multiple annotations a suitable 'ground truth'? Finally, AMBER also provides a mathematical formalism that can aid with comparing emotion representation schemes and analyse methods for converting between them. For example, if a Likert-type scale is used to gather annotated labels from multiple annotators, how would a numerical label and an ordinal label derived from that differ? What would they have in common? Is it suitable to use a clustering based approach is used to convert from a numerical label to a categorical label? These sort of questions have always been hard to answer due to the inherent complexity and ambiguity in affect. The AMBER framework is an attempt to overcome some of this difficulty by providing a mathematical language that explicitly articulates the core elements of all emotion representations.

\end{document}